\documentstyle[referee]{l-aa}
\begin{document}

\title{\bf On the True Energy Budget of GRB970508 and GRB971214}

\author{Abhas Mitra}

\institute{Theoretical Physics Division,  Bhabha Atomic Research Centre,
Mumbai-400085, India}

\thesaurus{07.35.1 ;07.36.1}

\maketitle

\begin{abstract}
We emphasize the already known idea that since GRB970508 released an energy of
$Q_\gamma\approx 10^{51} ~\delta\Omega$ ergs in soft gamma rays alone,
where $\delta \Omega$ is the solid angle of the beam,  the
actual energy of the  $e^+e^- p$ fireball driving the blast wave could be
considerably higher than this value, $Q_{FB} >Q_\gamma$.  We further argue
that, for reasonably large values of $\delta \Omega$, as is probably
suggested by the radio observations, the value of $Q_{FB} \sim 5.
10^{51}~t_m^3~n_1$ erg, for GRB970508; where $n_1$ is the number density
of the ambient medium in units of 1 proton/cm$^3$ and $t_m$ is the epoch
in months when the associated radio-blastwave degrades to become mildly
relativistic. Thus the value of $Q_{FB}$ for GRB970508 could be as large
as $10^{53}$erg or even much higher.  This idea is corroborated by
GRB971214 for which the value of $Q_\gamma \approx 2.
10^{52}~\delta\Omega$ is  much higher than the corresponding value for
GRB970508.  It is likely that GRB971214 has a correspondingly higher value
of $Q_{FB}$.  We discuss that it is unlikely that this energy, $Q_{FB}
\sim 10^{53}$erg was liberated by the central engine by a direct
electromagnetic mode. On the other hand, as conceived by several previous
authors and as is suggested by supernova theories, the $e^+e^- p$ fireball
(FB)
driving the blast wave is likely to be preceded by a much stronger
neutrino burst: $\nu +{\bar \nu}
\rightarrow e^+ +e^-$. Although,  the efficiency for $e^+ e^-$ production
by this latter route is usually found to be as low as $\eta_\pm \sim
10^{-3}$, we point out that for very high values of neutrino energy
release $Q_\nu >Q_{FB}$, it is possible that, the value of $\eta_\pm$
increses substantially.  By considering that, the value of $Q_{FB}$ for such
GRBs is indeed $\sim  10^{53}$ erg, we envisage  that the energy of the
actual neutrino burst/wind could be as high as $\sim 10^{54} -10^{55}$ erg
(here we ignore likely loss of energy by gravitational energy mode).

This energy might be had from general relativistic collapse of a massive stellar
core having {\em initial} gravitational mass, say, $M_i \sim few~ M_\odot$
to a stage having a potential well much deeper than what is present on a
canonican neutron star surface.  Accordingly, we predict that the new generation
large neutrino telescopes may detect neutrino burst of energy $Q_\nu \sim
10^{54} -10^{55}$ ergs, with neutrino energies touching $\sim 1$ GeV, in
coincidence with GRBs.

\end{abstract}


\section{INTRODUCTION}
The discovery of an absorption line with redshift $0.835<z<2.3$ in the
optical afterglow of the gamma ray burst (GRB) of 8 May, 1997 has now
decisively shown that atleast some of these events lie at cosmological
distances (Djorgovski et al. 1997, Metzger et al. 1997). The observed
fluence of $3. 10^{-6}$ ergs cm$^{-2}$ in the (20 keV- 1 MeV) band
(Kouveliotiou et al. 1997), suggests an energy release of $Q_\gamma\approx
10^{51} ~\delta\Omega$ ergs in soft gamma rays alone, where $\delta
\Omega$ is the solid angle of the beam.  Further, the discovery of a
redshift as high as $z=3.42$ (Kulkarni et al. 1998) for the highly
luminous event $GRB971214$ implies a value of $Q_\gamma \approx 2.
10^{52}~\delta\Omega$. Taken together, they confirm the view that
most of the classical GRB sources indeed lie at cosmological distances and are
accompanied by an amount of energy emission (in soft gamma rays alone)
which could be as large as 2-3 orders of magnitude higher than what,
so far, most of the existing models have attempted to explain, i.e.,
$Q_\gamma
\sim 10^{49-50}~ \delta~ \Omega$ erg. The objective of the present paper is
to consolidate the existing view (Waxman, Kulkarni \& Frail  1998, Kulkarni et al.
1998, Ramprakash et al. 1998) that the  observed energy liberation in the
soft gamma-band with such  energy liberation is preceded by a  larger
energy release $Q_{FB}$ in the form of a $e^+ e^- p$ fireball (FB). Then
we would like to enhance another known point that  it is extremely difficult
to conceive of any  central engine which can liberate the $e^+ e^- p$
fireball energy  $Q_{FB}$ by a direct electromagnetic process.  Therefore,
as suggested by most of the original cosmological GRB models, this FB is
most likely preceded by  an even much more energetic $\nu-\bar\nu$ burst.

We develop this paper initially around the observations of GRB970508,
because though it is much less energetic than GRB971214,  it is the only
event for which radio afterglow studies have revealed  that the associated
blast wave remains relatistic several months after the main event (Frail
et al. 1997, Waxman et al. 1998).  This suggests that the original kinetic
energy of the $e^+e^- p$ FB could be considerably higher than $Q_\gamma$.
If the radio observations could precisely fix the epoch when the blast
wave becomes marginally relativistic with a bulk Lorentz factor (LF)
$\Gamma_{NR}=1+
a < 2$, ( $a < 1$), it may be possible to to estimate the energy
of the blast wave (at this epoch).  The lab frame radius of the burst at
this stage would be (Waxman 1997)
\begin{equation}
R_{NR} \approx
4 \Gamma_{NR}^2 c t_{NR} \sim 3.2 . (a +1)^2 10^{17} t_{m}~{\rm cm}
\end{equation}
 where $c$ is the
speed of the light
and $t_{m}$ is the lab frame transition time in months.
If it is assumed that during this period the blastwave has been sweeping
the interstellar medium (ISM) of the host galaxy having a number density
of $n_1$ in units of 1 proton cm$^{-3}$ the terminal energy would be:
\begin{equation}
Q_{NR} \approx \left({\delta \Omega \over 3}\right) R_{NR}^3 m n_1 a
c^2 \approx  2. 10^{49} n_1 a (a +1)^6 t_{m}^3  \delta \Omega
~{\rm ergs}
\end{equation}
It is important here to note that the foregoing expression is actually
highly sensitive to the precise value of $a$ and $t_m$. And also
one must add $E_{\gamma}$ and all other
radiated forms of electromagnetic energy to the foregoing value of
$Q_{NR}$ in order to infer the original value of
the energy of the  $e^+ e^- p$ FB, $Q_{FB} =a_\gamma Q_\gamma$,
where $a_\gamma >1$.
Assuming that, for GRB970508, the blastwave was
indeed found to have $a\approx 1$ at $t_m \approx few$
 (Waxman et al. 1998), we would have
\begin{equation}
Q_{NR}
 \approx  1.3. 10^{51} n_1  t_{m}^3 ~ \delta \Omega~{\rm ergs}
\end{equation}
showing that, it is possible that $a_\gamma \gg 1$.
As the blastwave gets dergraded to non-relativistic regime, there may be a
rather sudden change in the (absolute) value of the exponet of the synchrotron
spectrum ($\alpha$). Obervational determination of such a fiducial point alone would
not be sufficient to fix the precise value of $t_m$ and what may be required is a
detailed pattern on the evolution of $\alpha$ preceding this epoch.
Obviously, one needs to
 estimate the value of $n_1$ too. We do
not know whether radio observations have actually yielded such detailed information.

The radio observations of GRB970508, have however clearly indicated that
the blastwave evolution is of adiabatic nature. Then, in principle, it is
possible to fix the value of $Q_{FB}$ in terms of a simple model which
involve the magnetic field equipartition parameter, $\xi_B$, and
electron-proton energy equipartition parameter, $\xi_e$. Since there is no
theory to independently estimate $\xi_B$ and $\xi_e$ (which could be
slowly evolving), we feel, it is really not possible to theoretically
predict the value of $Q_{FB}$ even if such models as such are quite satisfactory.
On the other hand, we feel, it would be more desirable to fix the value of
$Q_{FB}$ by means of painstaking study of the evolution of $\alpha$, and
this input should, in turn be used to estimate the values of $\xi_B$ and
$\xi_e$ subject to the uncertainty about the value of $n_1$. Probably, by
an iterative scheme involving appropriate theoretical models one may try
to obtain more reliable value for all the relevant quantities.

 Further, if, as suggested by the radio observations, if
$\delta
\Omega \approx \pi$ (Waxman et al.  1998), we would finally have
have
\begin{equation}
Q_{NR}
 \approx  4. 10^{51} n_1  t_{m}^3 ~{\rm ergs}
\end{equation}
The radio observation for GRB970508 might be consistent with a value of
$t_m \sim 4-5$ or even higher, and therefore, even if the value of $n_1$
turns out to be considerably smaller than unity, it is possible that
 $Q_{FB} \sim Q_{NR} > 10^{53}$ erg.

It could be so, because, if we assume a value of $\delta \Omega \sim \pi$
for GRB971214 too, the value of $Q_\gamma$ in this case would be $\approx
10^{53}$ erg. This burst being four times longer than GRB970508, it is
likely that a substantial portion of $Q_{FB}$ is already spent in
producing the main burst so that the eventual value of
$Q_{FB} \sim Q_\gamma \sim 10^{53}$ erg. We
will see below that any reasonable extension of the direct electromagnetic
modes of energy extraction is unlikely to yield this huge value of
$Q_{FB}$.

\section{Pulsar Mode}
If, following Usov (1992) and Duncan \& Thompson (1992), we assume that
some pulsars are born with period, $P\sim 1$ms,  the
initial Rotational Kinetic Energy (RKE) of such pulsars would be
\begin{equation}
RKE ={1\over2} I \Omega^2 =5. 10^{52} ~I_{45} \Omega_4^2~{\rm erg}
\end{equation}
where $I_{45}$ is the moment of inertia of the NS in units of $10^{45}$ g
cm$^2$ and $\Omega_4$ is the cyclic frequency in units of $10^4$.
 It follows that, the
magnetic dipole luminosity could be as large as
\begin{equation}
L_{em} \approx 2. 10^{50} B_{15}^2 \Omega_4^4 {\rm erg~ s}^{-1}
\end{equation}
where $B_{15}$ is the surface magnetic field in units of $10^{15}$G.
The time scale for electromagnetic mode is thus
\begin{equation}
t_{em}= {RKE\over L_{em}} \approx 150 B_{15}^{-2} ~\Omega_4^{-2}~{\rm s}
\end{equation}
It is well known that rapidly spinning pulsars may radiate more of
gravitational radiation than electromagnetic radiation (Ostriker \& Gunn
1969) because of equatorial ellipticity ($\epsilon$).  For a pulsar with a
solid crust, the major source of $\epsilon$ could be the dynamical
anisotropies associated with the magnetic field. On the other hand, note
that, a freshly born pulsar is believed to be very hot with core temperature
well above $10$ MeV. This means that the crust will be liquid (Meszaros \&
Rees 1992) and the solid crust must be formed when they become
sufficiently cool by intense $\nu-{\bar \nu}$ radiation to temperature,
probably, below, 1 MeV. Since the whole NS behaves like a self-gravitating
liquid at this stage, a large RKE is likely to introduce an equatorial
ellipticity far exceeding what is what is obtained by considering only the
magnetic field as the source of such ellipticity.

First note that, the mean density of a canonical NS with mass $1.4
M_\odot$ and radius $10$ Km is $\bar \rho \approx 7. 10^{14}$ g cm$^{-3}$.
Then the critical cyclic frequency at which mass sdedding occurs is
$\Omega_k \approx 0.66 (\pi G \bar\rho)^{1/2} \approx 0.7. 10^4$ Hz,
which corresponds to a critical period of $P_k \approx 1$ ms. In our
attempt to first understand the problem within the purely classical paradigm,
 we ignore
the complexities and uncertainties associated with the actual equation of
state (EOS) of hot and liquid NS (for a {\em cold} NS with solid crust, the cold
 EOS permits value of $P_{min} \sim 0.3 -0.5$ ms). A value of $\Omega_4
\ge 1$, thus may not at all be allowed for the liquid like hot NS.
If  the actual value of $\Omega< 0.2 (\pi G \bar\rho)^{1/2}\approx 0.33
\Omega_k\sim 2. 10^3$Hz, the  rotating configuration will be described by Maclaurian
Spheroids with the two equatorial principal axes $a=b \neq c$, where $c$
(not to be confuse with speed of light) length of the principal axis along
the rotation vector (Chandrasekhar 1969). Equatorial eccentricity being
zero, there will not be any gravitational radiation in such a case.

However, since in the present problem, we are working in a region $\Omega
\rightarrow \Omega_k$ (actually by considering $\Omega_4 \approx 1$, one,
unphysically overshoots $\Omega_k$ barrier), {\em we are not necessarily dealing with
axisymmetric configurations} because for $\Omega >0.33 \Omega_k$, the
Maclaurian spheriods are {\em dynamically unstable} and tends to
degenerate into Jacobi Ellipsoids with $a \neq b\neq c$ (Chandrasekhar
1969 and ref. therein) with equatorial ellipticity
\begin{equation}
\epsilon = {2 (a-b)\over (a+b)}
\end{equation}
 Along the Jacobi sequence, $\epsilon$ increases
monotonically in keeping with the increasing  angular momentum. In
fact, Poincare (see Chandrasekhar 1969) showed that, the Jacobian sequence
eventually bifurcates into a new sequence of {\em pear shaped} configuration where
\begin{equation}
{b\over a}= 0.432232; \qquad {c\over a}=0.345069; \qquad \Omega=0.28403 \pi G \bar\rho
\end{equation}
In this above limit the value of $\epsilon \rightarrow 0.8$.
 Consequently, even when, we exclude the regime of extreme ellipticity by
hand, we find that such hot and fluid-like ultrafast pulsars may emit
superstrong gravitational radiation with a luminosity (Ostriker \&  Gunn 1969):
\begin{equation}
L_{gw}= {32\over 5} {G \epsilon^2 I^2 \Omega^6 \over c^5} \approx 1.5. 10^{53}
I_{45}^2 \Omega_4^{6} \epsilon_{.1}^2 ~{\rm ergs~ s}^{-1}
\end{equation}
where $\epsilon =10 \epsilon_{.1}$. It is trivial to see that the
 corresponding {\em
instantaneous} time scale for emission of gravitational radiation would be
\begin{equation}
t_{gw} = {RKE \over L_{gw}} \approx 0.2 I_{45} \Omega_4^{-4} \epsilon_{.1}^{-2}
 ~{\rm s}
\end{equation}

 Since $t_{gw} \ll t_{em}$, the NS would primarily emit its RKE  by
gravitational radiation within a time $\sim 0.1$s rather than by any
significant relativistic wind, and acquire a value of $P >3 P_k\approx
3$ ms for
which the gravitational radiation mode will be quenched. The consequent
value of  $L_{em}$  would also be insignificant for the
requirement of GRB problem once $P$ degrades to this range.

It may be relevant to point out that in a
recent work extending the Usov-type mechanism (Blackman \& Yi 1998), the
putative hot ms pulsars have been classified into [1] Supercritical strong
field rotator (SPS) for $\Omega >\Omega_{crit}$, ($\Omega_{crit}$ being the
critical luminosity over which pulsar spin down is dominated by emission
of gravitational radiation) and [2] Subcritical strong field rotators
(SBS) with
$\Omega <\Omega_{crit}$. For the SPS, the initial spindown is dominated by
strong gravitational radiation with e-folding $t_{gw} <1$s. This is
somewhat analogous to our conclusion reached above. It is only with reference
to the
latter class, the SBS, with a relatively lower value of $L_{em}$, one may
endeavour to correlate observation of long GRBs. But, it is seen that, the
peak luminosity in such cases would be well below $10^{50}$ erg/s, which is
grossly insufficient for GRB970508 and 971214 ($\delta \Omega \sim \pi$).

Also, recently, in several important theoretical works
 it has been pointed out that (Andersson, Kokkotas, \& Schutz
1998, Lindblom, Owen \& Morsink 1998) even when the value of
$\Omega < 0.33 \Omega_k$, or even if the above crude discussion  on
probable emission of gravitational radiation from the rotating Jacobi
Ellipsoids were not precise, the NS may spin down by emitting
gravitational radiation. When the NS is sufficiently hotter than $\sim
10^9$ K, and one would consider
the overall configuration of the fluid to be broadly spheriodal,
some degree of non-axisymmetry will set in because of velocity and density
perturbations (r-mode). Lindblom et al (1998) estimate that because of
r-mode instability, the NS would spin sown to a period $P_{cool} \sim 13
P_k$ by the joint action of neutrino viscosity and gravitational radiation
within a neutrino emission dominated cooling time, $t_{cool}$, to a
temperature $\sim 10^9$ K. However, this paper used a somewhat old formula for NS
cooling which gives a very large value of $t_{cool} \sim 1$ yr.  On the other
hand, the  more recent work of Andersson et al.  (1998), which
claims to be more accurate than the previous one, points out that, if one uses more recent
work on NS cooling by direct URCA process (Lattimer et al.  1994), the
value of $t_{cool}$ could be as small as $\sim 20$ s. It is clear that, in
this case, because of rapid decrease in the value of $\Omega$, the value
of $L_{em}$ would again be insignificant.

Even in the former case of a supposed very high value of $t_{cool}$, the
value of $L_{em}$ may be much less than what is implied by eq. (6). This
is because of the fact that, in the theories of formation of superstrong
NS magnetic field, the field is not generated
spontaneously and automatically at the moment of birth of the nascent proto-NS.
Instead, it is believed to be generated either by differential
rotation (Kluzniak \& Ruderman 1997) or by dynamo amplification (Duncan \&
Thompson 1992). The latter process results in a faster generation of
magnetic field and is presumably caused by vigourous convection {\em
driven by strong neutrino flux} $F_\nu \sim 10^{39}$ erg/s/cm$^2$. One may
have such a strong neutrino flux only if $t_{cool} \sim 10$s. If the
neutrino cooling is too slow, $t_{cool} \sim 1$ yr, the value of $F_\nu$
will accordingly be too low, and {\em
no strong magnetic field may at all be generated}. In this case, the Usov type
models will be irrelevant.

 Consequently, assuming a NS is born with a $\sim 1$ms-period, it spins
down to 10-20 ms range by emitting an energy $>10^{52}$ erg in the form of
gravitational wave energy and probably much more energy in the form of
neutrinos.  Thus, because of the r-mode instability a supposed ultrfast
pulsar will hardly have an
opportunity to acquire a large $L_{em}$ independent of the actual value of
$t_{cool}$.

Note that, in contrast, an old and recycled pulsar with spin $\sim 1$ms
will be very cold with a thick and solid crust whose value of $\epsilon$
may indeed be derermined by $B$ and could be very low. Such cold stars
with $T <10^9$ will have superfluid interior, where the r-mode gets completely
supressed (Andersson et al. 1998, Lindblom et al. 1998).

\section{Accretion Disk Mode}
A NS-NS or a NS-Black Hole (BH) collision may lead to the formation of a
transient compact object-disk
type configuration (Narayanan, Paczynski, \& Piran 1992, Ruffert et al.
1997, Woosley 1993), and following the previous ideas by several authors,
(Lovelace 1976, Blandford \& Znajek 1977)
 that such configurations may  generate
relativistic beams by electromagnetic process, it has been recently
proposed that NS-NS or NS-BH collisions may may generate superstrong
relativistic winds (Katz 1997, Rees \& Meszaros 1997). In very broad terms the power radiated by such
scenarios could be understood in terms of the Lovelace (1976) model
by assuming the disk to be  Keplerian and thin. The magnitude of the
induced disk electric field would be  (standard polar coordinates are used):
\begin{equation}
E_r \sim  {1\over c} v_\phi B_z
\end{equation}
The corresponding voltage drop between the inner radius $r_i$ and the
outer radius $r_o$ would be (Lovelace 1976):
\begin{equation}
V_{io} \sim {1\over c} \int_{r_i}^{r_o} dr~ v_\phi B_z
\end{equation}
where $r_i$ is the inner radius and $r_o$ is the outer radius of the disk.
In the absence of a any reasonable information about the profile of $B_z$,
one may be forced to consider the most favourable order of magnitude
estimate of the foregoing equation by considering $B_z =B_z (r_i)=B_i$,
$v_\phi =v_\phi (r_i)=v_i$, and $r=r_i$
\begin{equation}
V_{io} \sim {v_i\over c}  B_i r_i ~{\rm cgs} \sim 10^{22}
{B_{4}} M_{10}~ {\rm V}
\end{equation}
where $B_{x}= (B_i/ 10^x {\rm G})$, $M_{x}= (M / 10^{x} M_{\odot})$, and
we have taken $r_i$ as the radius of the last
stable circular orbit and $v_i$ as the corresponding Keperian value:
\begin{equation}
 r_i= 3 R_g = 6 G M/c^2; \qquad v_i \sim \left({G M\over r_i}\right)^{1/2} = {c \over \sqrt{6}}
\end{equation}
 The
current flowing along the disk axis would be $I \sim c V_{io}$ and, using
eq.(14),  the
maximum beam power output would be
\begin{equation}
L_b \sim I V_{io} \sim c V_{io}^2 \sim c {B_i^2 r_i^2\over 6} \sim
5. 10^{49} {\rm ergs~ s}^{-1} {B_{4}}^2 M_{10}^2
\end{equation}
 For a (stellar mass) NS with weak magnetic fields,
 the disk may nearly touch
the surface and, the above value of $r_i$ would be appropriate for such
cases too. If the above model is extended for such a stellar mass NS, we
would obtain a value of $L_b$ which is quite high for X-ray binaries:
\begin{equation}
L_b \sim 10^{38}~ B_{8}^2 ~{\rm erg~s}^{-1}
\end{equation}
where the value of $B_{8} ={B_i/ 10^8 ~{\rm G}} \sim 1$
 corresponds to either the disk inner
edge touching the NS-surface or the case of a closeby inner edge immersed in
the magnetosphere of the NS. This kind of NS-disk model was actually
suggested to explain the origin of supposed ultra high energy gamma rays
and cosmic rays from the X-ray binary Cygnus X-3 (Chanmungam \& Brecher 1985).
It was argued, in the eighties, that, Cyg X-3 might contain a rapidly
spinning NS, and therefore, either by direct pulsar action or by NS-disk
symbiosis, as has been proposed for the present GRB case, should be a
strong source of ultra high energy gamma rays and cosmic rays. By
overstreching the models, it was even envisaged that a single Cyg X-3 type
source could solve the problem of origin of ultra high energy cosmic rays
for the entire galaxy. However, it became, clear later, that such
theoretical precdictions were completely unfounded.
Further, if one has to fit such models into the cosmological GRB scenario,
one has to further overstretch such (unsuccessful) models by many
orders of magnitude; and
one would be compelled to assume an enhanced value of $B_i >10^{14}$G to obtain
a notional value of $L_b >10^{50}$ erg s$^{-1}$.

\subsection{Available Energy}
In case we are considering a BH-disk case, let us estimate first, what
could be the order of magnitude of the central BH. From eq.(15), we find the
period associated with the last circular orbit to be
\begin{equation}
P_i ={2 \pi r_i \over v_i} \approx 5. 10^{-4} ~\left({M\over {M_\odot}}\right)
{\rm s}
\end{equation}
Thus in order to explain sub-ms time structure in some of the GRBs (Bhat
et al. 1992), we
should consider a value of $M$ not exceeding a few solar masses (Woosley
1993).   For a spinning canonical NS, the value of RKE is limited by the
binding energy (BE), $\sim
3.  10^{53}$ erg. We note here that for a NS, there is a natural
electromagnetic coupling between the disk and the star, and, in principle,
it is possible that the supposed beam is partially fed by RKE too.
However, it does not at all mean that this is necessarily the case, and,
in any case, it does not mean that the entire RKE of the NS is tapped
during the finite life time of the disk. If the disk could be considered a
permanent structure, a rigid conductor, and rotating synchronously in
tandem with the spin of the NS, maintaining a perfect electromagnetic
coupling to the NS, it is possible in principle that, given very long
time, the entire RKE is harnessed.  But the resultant value of $L_b$ in
such a case would be determined by the value of surface magnetic field of
the NS, and not by the supposed enhanced value of $B_i >10^{14}$G. In
other words, in such cases, one reverts back to the pulsar (with intrinsic
dipole moment) models. Thus, in the model of Lovelace (1976) and Chamungum
\& Brecher (1985), the actual source of $L_b$ is the accretion power and
not the RKE of the central compact object. In fact the values of $L_b$
indicated by either eq.(16)  or eq.(17), roughly suggest the conversion
of an Eddington limited accretion power to electromagnetic power.

This conclusion is even more appropriate for the BH-disk case even though,
technically, for a maximally rotating Kerr-Newman BH, the effective value
of RKE $\sim 29\% ~M c^2$. This is so
because, unlike  a NS-disk case, a Kerr BH has no intrinsic charge, $Q=0$,
or magnetic moment, $\mu=0$. Thus {\em there is no natural
electromagnetic coupling between the BH and its disk}.  There are,
however, some theoretical estimates, based on {\em purely vacuum}
electrodynamics, that a stationary axisymmetric BH, placed in an external
uniform magneic field,$B_0$ may acquire  charge by the accretion process
whose asypmtotic value would be $Q=2 B_0  M$ (Wald 1974). If so, it might
be possible to harness the rotational energy of the BH, which, for a
maximal case could be $\approx 2. 10^{54}$ erg (taking a value of $M=3
M_{\odot}$).  Similar idea is also expressed by Blandford \& Znajek (1977).
Given a sufficiently long lived accretion disk (fed by external material),
such ideas might be relevant for jets observed in AGNs. Nonetheless, note
that, in a strictly axisymmetric  and steady case, the accretion process
would not deliver any net torque from the BH (Ruffini 1978) though it
might be possible that in a transient case some net torque may be derived from
the BH. Also note that the BH can acquire a substantial dipole moment only
after substantial accretion has taken place.  Thus, note that, even in the
AGN context, the source of jet power is believed to be primarily
the accretion power rather than the stored RKE of the BH (Begelman,
Blandford, \& Rees 1984).

In the transient GRB case, occurrence of substantial accretion
may imply the vanishing of the disk.
And unlike the case of a pulsar, the RKE of the BH (with no intrinsic
diploe moment),  even if it is substantial, can not deliver any
power once the disk has vanished.
 And for the transient GRB event, a very high value of $L_b$ can
be justified by assuming the release of accretion power within a few seconds.
 Thus, atleast for the GRB problem, the supposed high RKE of the BH
can not be meaningfully harnessed and the
maximum extractable energy available by all probable processes is the the
total BE of the disk of mass $M_d$ in the gravitational potential of $M$.
 In order that the
electrodynamic jet mechanism is successful, i.e. the potential drop
$V_{io}$ across is maintained, the insulating effect of the magnetic field
must be effective. This condition is ensured, if for most of the region of
the beam, we have $v \ll c$  (Lovelace, MacAuslan and Burns  1979), which
independently demands that $E_{beam} \ll M_{disk} c^2$.
For a NS-NS collision case the disk may have a mass of $\sim 0.1
M_{\odot}$ (Ruffert et al. 1997), and thus
the value of
\begin{equation}
E_{b} <E_{disk} \sim {G M M_d\over 2 r_i} = {M_d c^2\over 12}
 \sim 2.5. 10^{52} {\rm ergs}\left({M_d \over 0.1 M_{\odot}}\right)
\end{equation}
is insufficient for explaining
the energy budget of GRB970508.  However, assuming a case of NS-BH collision with $M_{disk}
\sim 1 M_{\odot}$ (Meszaros \& Rees 1997), it may  be  possible to
satisfy this bare energy requirement ($Q_{FB}$).

\section{Super Strong  Magnetic Field}
Let us also ponder how far we are justified in assuming a superstrong
$B\sim 10^{15}$G for a very hot new born pulsar or accretion torus. As to
the origin of such magnetic fields in young pulsars, Duncan \& Thompson
(1992) suggest that the convective turbulence associated with strong
neutrino transport with $F_\nu \sim 10^{39}$ erg/s/cm$^2$ causes such
magnetic field by the dynamo action. This tends to suggest that the strong
magnetic field is an aftermath of strong neutrino cooling and it need not
be present in the nascent and hot NS.  In a normal astrophysical or
laboratory plasma, microscopic turbulence may convert part of the bulk
kinetic energy into magnetic field. However, at the same time it should be
remembered that turbulence on macroscopically significant scales may have
a tendency to destroy the pre-existing strong magnetic field by producing
spatially and temporally incoherent current systems and by means of ohmic
heating. Thus, there may be an optimal level of turbulence conducive to
production of strong magnetic field and beyond which turbulence may be counter
productive.

In general, linear extrapolation of magnetic field generation ideas (in
relativistically  hot plasma), which are originally meant to explain much
lower fields, may not be proper because note that there is a natural
quantum unit of $B=B_q =m_e^2 c^3 /e \hbar \approx 4.4.  10^{13}~ {\rm G}$
for $B$. Here $e$ is the electronic charge and $m_e$ is the electron mass.
For $B= B_q$, the {\em Larmour radius of the electrons, whose current
ultimately gives rise to $B$, becomes equal to the electron Compton
wavelength}.  Thus it may be possible that one can realize a value of $B$
comparable to $B_q$ only when there is some degree of {\em macroscopic
quantum behaviour} of the underlying medium. Atleast, in cold neutron
stars this is the case in that the source of the magnetic field can be
ascribed to the existence of strong internal current systems in the form of
quantized flux tubes or fluxoids carrying elementary flux $\phi_0 =\pi
\hbar c/e$ (Bhattacharya \& Srinivasan 1995, and ref. therein).
The existence of strong internal currents may be inextricably
linked to the existence of a superconducting NS -interior (protons and
neutrons behave like superconducting medium in which the electron current
flows). The fluxoids have the {\em same sense of current} and $B\sim
N_F \phi_0$, where $N_F$ is the number of fluxoids threading each cm$^2$.
Such type II (or any other) superconductivity is a manifestation of
macroscopic quantum behaviour and can be operative only below certain
critical temperature, which, for NS interiors, is $T_c < 1 MeV$.
Therefore, it might be possible that, a new born NS, which must be very
hot does not possess the observed strong field, and, the strong field is
set up later at the expense of the turbulent and internal energy as the
star cools below $T_c$ and becomes quantum mechanically organized.

On the other hand, the catastrophic
NS-NS collision process, which is far from the spherically symmetric
near-adiabatic core-collapse scenario and results in a central compact
object of $T\sim 50$ MeV (Ruffert et al 1997), might destroy the preexisting
order and the magnetic field.  The same is even much more true for the
resultant disk which is very hot ($\sim 10$MeV) and macroscopically turbulent.
Recall here that, there is a critical
temperature above which the NS core or the disk ceases to be a
superconductor. Earlier theoretical estimate was that in the density range
of $T_c \sim 10^{13} - 10^{15}$g cm$^{-3}$, the transition temperature lies
between $\sim 1-20$MeV. But more recent and and refined estimates find $T_c
\sim 0.5$MeV even at the highest densities (Bhattacharya \& Srinivasan 1995).
Thus, the hot compact object
or the disk resulting from tidal distortions is most unlikely to be the site
of a magnetic field whose value exceeds the characteristic quantum value
$4.4. 10^{13}$G.  Probably,  the modest macroscopic
quantum behaviour implied by a type II superconductor can not explain a
$B>10^{13}$G, the maximum value for radio pulsars, and it may require a greater
degree of quanum coherence to have a field stronger than this. If one
reqires a value of $B \gg B_q$, it may be more logical to conceive that
superconductivity is due to flow of protons rather than of electrons. This
will, however, demand, quantum coherence on much larger scale.

  It is thus particularly difficult to conceive how the hot, turbulent,
and incoherent accretion disk resulting from a NS-NS or NS-BH collision
may possess a value of $B\sim 10^{14}~{\rm G} > B_q$.

At any rate the {\em sustenance} of a strong NS magnetic field is
certainly not due to any strong macroscopic turbulence, because, in a cold NS
core or crust, there is hardly any macroscopic turbulence. It must be
mentioned now that recently, in a remarkable observation, the so called Soft
Gamma Ray Repeaters (SGRs) have been identified to be associated with a
class of NS with strong magnetic field $\sim $few $10^{14}$ G (Kouveliotou
et al. 1998). These class of NSs
with superstrong magnetic fields have been termed as ``magnetars''. But,
from our view point, the important point is that the magnetars are
{\em extremely slow rotators} with $P \sim$ few s. Their spin down luminosity
$L_{em}$ is very low having a value of only $\sim 10^{34}$ erg/s. The
stored magnetic energy of such stars is in the range of $\sim 10^{46-47}$ erg,
which far exceeds the RKE $\sim 10^{44}$ erg (at the present epoch).
 It is
interesting to note that the X-ray and particle emissions from the
magnetars, which may  fuel the SGR activity, are powered not by rotation but by
the sporadic release of the stored magnetic energy. The magnetars
discovered now  may be several thousand years old with temperature,
certainly, much below 1 MeV.

\section{Electrodynamics Immersed in Accretion}
A general requirement for realizing the vacuum potential drop $V_{io}$
is that the density of the ambient plasma is not exceedingly
larger than the characteristics Goldreich-Julian density (Goldreich \&
Julian 1969) :
\begin{equation}
n_{GJ} \sim  10^{-2} {B_z \Omega} ~{\rm cm}^{-3}
\end{equation}
where $\Omega$ is the cyclic frequency of the accelerator in Hz.
This is the density of space charge that arises in the magnetosphere of an
unipolar inductor type device in the absence of any externally injected
plasma. It is because of the unavoidable presence of this space charge
that the
actual available potential drop in the magnetosphere becomes considerably
smaller than the vacuum value $V_{io}$. And one of the fundamental and
pending problem of pulsar type mechanism is to self-consistently evaluate
the value of the modified lower voltage $V_{io}' <<V_{io}$. If it were not
not so, many of the X-ray binaries containing either rapidly spinning NSs
or modestly spinning white dwarfs would be strong soures of ultra-high
energy cosmic rays. Even for isolated radio pulsars like Crab and Vela,
the absence of emission of either ultra-high energy cosmic rays and gamma
rays (the observed gamma rays lie in the TeV-range) would suggest that,
even in the ideal cases, it is difficult to realize the vacuum
potential drop $V_{io}$. Further, when
external plasma is present, the problem becomes even poorly defined, and,
all estimates based on the vacuum value of $V_{io}$ become highly suspect.
In particular, if the external plasma density $n \gg N_{GJ}$, at a certain
stage, the pulsar type mechanism ceases to operate. This is corroborated
by noting the fact that althogh most of the bright X-ray binaries contain
NSs with sufficiently high magnetic field, in general, they do not operate
in the {\em pulsar mode}, i.e., they undergo spin up rather than spin down
process. Also, intermittently, during stages of low accretion rate,
usually considerably lower than the Eddington rate, some of the pulsars in
the X-ray binaries, undergo brief spells of spin-down.

For  the  the innermost circular
orbit, we have
\begin{equation}
\Omega_i \sim \left(GM /r_i^3\right)^{1/2} = c^3 /(6^{3/2} G M)
\end{equation}
 implying a value of
\begin{equation}
n_{GJ} \sim \left({10^{-2} B_i c^3 \over 6^{3/2} G M}\right)
\sim 2. 10^2 B_i \left({M\over M_{\odot}}\right)^{-1}
\end{equation}
 On the other hand, for accretion at the Eddington rate, ${\dot M}_{ed} \sim
10^{18}$g s$^{-1}$, in the approximation of spherical accretion, by using
eq.(18), we find that, the
density of the accretion plasma near the innermost region would be
\begin{equation}
n_{ed}= {{\dot M}_{ed} \over 4\pi r_i^2 v_i m_p}\sim 5. 10^{18}
\left({M\over M_{\odot}}\right)^{-2}~{\rm cm}^{-3}
\end{equation}
Further, we define a ratio to quantify the degree of contamination of the
extraneous plasma:
\begin{equation}
\xi_{em} = {n_{ed}\over n_{GJ}} \sim 3. 10^{16} B_i^{-1}~ \left({M\over M_{\odot}}\right)^{-1}
\end{equation}
For a supermassive BH of $M\sim 10^8 M_{\odot}$ having a value of $B_i
\sim 10^4$G, or for an accreting millisecond binary pulsar with a low $B_i
\sim 10^8$G, and whose
accretion disk may be almost touching it, we find a value of $\xi_{em}
\sim 10^4-10^5$. Note that, even if we consider the disk to be ideal and thin,
there is alwaya a corona above and below the disk. Further,
in all realistic cases, in particular, in case of minor deviation from
exact axisymmertic accretion geometry, there may be a small component of
quasi-spherical flow with density $n_{sp}\sim \zeta \xi_{em} n_{ed}$. This
is more true for a thick accretion disks or tori. Such extremely weak
quasi-spherical flows or corona need not inhibit the formation of effectively baryon
free funnels along the symmetry axis, however they can certainly quench
the vacuum potential drop estimated in eq.(16).

Although, most of the
accretion power of AGNs is primarily manifest in the form of ultraviolet
or X-rays, we note that relativistic bipolar flows with luminositities,
sometimes, comparable to $L_{accretion}$ seems to be a common features in
active galactic nuclei. We also note that radiation driven or hydrodynamic
jet models have difficulty in achieving a value of $\Gamma\ge 2$,
(Begelman 1994)  whereas most of the superluminal flows as well as
gamma-ray observations require a value of $\Gamma < 10$. Here we may add
that for the recently observed galactic `micro quasars', in some cases,
the inferred value of $\Gamma \sim 2.5$ while in one case, the value of
$\Gamma \sim 10$ (Mirabel \& Rodriguez 1997, Hjellming \& Rupen 1995).
 And though for radiation driven jets, it is generally
found difficult to obtain values of $\Gamma >2$, recently, by using the
socalled ``Compton Rocket Effect'' Reanaud \& Henri (1997) have attempted
to show that radiation
driven disk-jet mechanisms can overcome such limitations, and, in fact,
attain a value  of $\Gamma \sim 60$ in the AGN context. However, the value
of the terminal Lorentz factor, in this model, is given by
\begin{equation}
\Gamma \sim 2 \epsilon_0^{-1/3}
\end{equation}
where $\epsilon_0$ is the mean energy of the disk photons in units of $m_e
c^2$
For the AGN problem, the value of $\epsilon_0$ is in the ultraviolet
range, but for the GRB problem, $\epsilon_0 \sim 1$, and therefore, again,
we have $\Gamma \le 2$.

Thus in the absence of alternative more successful theories about such
relativistic flows (Begelman 1994), in a rather generous manner, let us
assume here that it is the magnetically dominated disk-BH-jets which are
working in the AGNs.  By considering $L_b \sim L_{ed} \sim 10^{46} M_8$
ergs s$^{-1}$, we estimate, $B_{i4} \sim 1$ for $M_8
\sim 1$ so that typical $\xi_{em}^{AGN} \sim 3. 10^4$ and $\zeta <
10^{-4}$.

 In contrast, note that,  except for the recently discovered galactic
superluminal sources mentioned above, most of the known bright X-ray binaries (with
accretion disks) do primarily emit the accretion power in X-rays and not
by relativistic beams. The radio-jets observed in many X-ray binaries like
Cyg X-3, SS433, Sco X-1, are non-ralativistic with luminosities
insignificant compared to Eddington luminosities. And in the light of the
recent optimistic work on radiation driven jets (Reanaud and Henri 1997), it is entirely
possible that, purely electromagnetic jet mechanism is neither necessary
nor functional in the X-ray binaries. Even, if one assumes
that, for the stellar mass cases too, $L_b \sim L_{ed} \sim 10^{38}$ ergs
$^{-1}$, one would obtain a maximum effective value of $B_i \sim 10^9$G
with a corresponding value of $\xi_{em}^x \sim 10^9$. In the framework of
this probable relativistic jet-mediated liberation of accretion power, we
may explain the general absence of such activity in the X-ray binaries by
restricting
$\zeta \gg 10^{-9}$. Recall again that there will be extraneous plasma
 also be in the form an
accretion disk
corona and whose density could be proportional to the accretion rate.

The GRB disk case is similar to  a stellar mass X-ray binary case with the
difference that in the former case, we assume need a much higher value of
$B_i \sim 10^{15}$G. But, on the other hand, note that for the GRB case,
assuming that $0.1 M_{\odot}$ is accreted in $\sim 10$s, we  will have an
accretion rate of ${\dot M} \sim 10^{31}$g s$^{-1} \sim 10^{13}~{\dot
M}_{ed}$, so that
\begin{equation}
\xi_{em}^{GRB} \sim 10^{30}  B_i^{-1} \left(M/ M_{\odot}\right)^{-1}
\end{equation}
Thus, even for an assumed high value of $B_i \sim 10^{15}$G for a
GRB-disk, we will have $\xi_{em}^{GRB} \sim 10^{15}$, and with a value of
$\zeta \gg 10^{-9}$, the density of the quasi-spherical flow $n_{sp} \gg
n_{GJ}$, probably $n_{sp} \sim 10^{11} n_{GJ}$. Also note that, while
we observe the AGN or X-ray binaries in a quasi-steady state, probably,
atlest thousands of years after their formation, in the GRB disk case,
what is really formed after the catastrophic collison is a torus or simply
a
cloud (Ruffert et al. 1997) which may be settled into a more-steady
torus. And for such cases, even if the accretion were limited by the
Eddington rate, the chance of a minor quasi-spherical accretion would be a
genuine difficulty in achieving any idealized electromagnetic accretion
machine. Thus it is very difficult to see how electromagnetic modes of
energy extraction based on the idea of accelerators immersed in vacuum can
work for the GRB case.  In fact, the work of Ruffert et al. (1997) shows
that, it is indeed likely that most of the accretion power associated with
the GRB-disk is used in producing neutrinos.

\section{Summary \& Discussion}
In the past, purely electromagnetic modes of GRB origin, mostly involving
(isolated) pulsar magnetospheres, have been considered from time to time
in the context of galactic models or at best for the so-called
galactic-halo models. Energy constraints discouraged extension of such
models for the cosmological perspective because of the following simple reason.
The radio pulsars are
probably the best studied high energy astrophysics objects and there are
strong observational reasons to believe that the pulsar magnetic fields do
not exceed $10^{13}$G and pulsars are not born with a period shorter that
$\sim 10$ms (Bhattacharya \& van den Heuvel 1991). While, the constraint
on the initial spin period persists, recent observations have found that,
there are a class of pulsars with superstrong magnetic fields of few
$10^{14}$ G (Kouvelotou et al. 1998). However, given their age of few
thousand years,
they are very slow rotators with $P
\sim$ few s compared to the X-ray pulsars or normal radio pulsars. The quiesent electromagnetic activity of such pulsars are
completely insignificant for any kind of GRB activity, however, the
intermittent release of the stored magnetic energy, plausibly fuels the
SGR activity. As far has classical GRBs are concerned, these, magnetars
may not be relevant.

In fact, recently, Spruit \& Phinney (1998) have suggested that pulsars
might be born as slow rotators because   the magnetic locking between the
core and the envelope of the progenitor prevents the core from spinning
rapidly. This suggestion may be generally true
although there must be exceptions like Crab for which the initial spin has
been estimated to be $\sim 19$ ms (Glendenning 1996). In particular,
Spruit \& Phinney envisage that magnetars are born as slow rotators with
$P > 2$ s.

On the other hand, the elaborate
realistic numerical computations (though primarily Newtonian) to study the
physics of NS-NS collisions (Ruffert et al. 1997) showed that the value
of $Q_{FB}$  could be as low as $10^{47} -10^{48}$ erg, which
falls short of the {\em previously assumed
requisite value} $Q_{FB} \sim 10^{51}$ ergs
by several orders of magnitude.  In view of such discouraging results
(from the GRB view point), it
became, probably, important and necessary to reconsider the
electromagnetic models for the cosmological GRBs  too.  However, in this
paper, we probed some of the consequences for extending the normal
electromagnetic models involving stellar mass compact objects with $L_b <
10^{38}$ erg/s to scenarios with intended $L_b \sim 10^{50-51}$ erg/s or
even higher.

A hot and fluid like NS with a assumed value of $\Omega \sim 10^4$, as was
{\em believed to be required for the GRB problem in the pre-GRB971214
era}, is likely to lose practically the entire initial RKE by
gravitational radiation because of the equatorial ellipticity associated
with its Jacobi Ellipsoid configuration. It is then expected to settle to
a much lower value of $\Omega <0.33 \Omega_k$ within $<1$s of its birth.
Even then, it will be subject to r-mode instability. If the intial
$\nu$-cooling is rapid enough, $t_{cool} \sim 10$s, neutrino flux driven
conduction may set up a super strong magnetic field; but, by this time,
the bulk viscosity and gravitational radiation
may bring down the spin to $\sim 10 -20$ ms (we have already mentioned
that such strong field magnetars might actually be born with $P >2$ s).
Thus, these models will fail to explain GRBs.
And with an actual value of $Q_{FB} \sim 10^{53}$ erg, such models become
even much more fragile.

We also recall here that the approximate formation time of a NS out of the
proto-NS is determined by the fairly long neutrino diffusion time scale of
$\sim 10$s (Shapiro \& Teukolsky 1983) and this has been confirmed by SN1987A. Since this time scale is
comparable to or, in fact larger than, most of the GRB time scales, it is
unjustified to consider the sudden birth of a NS (of high magnetic field)
in isolation and then  seperately study its its probable spin-down or
cooling on a time scale of few seconds. Such supposed prompt spin down has
to be studied as an integral part of the final stages of the preceding
collapse process.  Following the discussion of section 1., as soon as, the
proto-NS would be envisaged to acquire high spin, the associated
gravitational damping would constrain the value of $P >3 P_k \sim 3$ms or
to a much higher value. In fact, the formation stages of the torus would
also certainly involve wobbling and spinning of the would be torus
material. And this would again mean huge loss of angular momentum and
kinetic energy from the system, though, it would practically be impossible
to estimate such losses. The formation of NS always involve release of its
BE $\sim 3. 10^{53}$ ergs primarily in the form of neutrinos and antineutrinos.
And this may indeed entail strong neutrino driven convection, and, yet it
is certain that most of the neutron stars, except the few magnetars, are
not born with super strong magnetic field. So our understanding of genesis
of super strong NS is far from complete and we need to invoke the idea of
such strong fields with caution.

  Recently, Kluzniak \& Ruderman (1997) have
considered a scenario in which the new born pulsar has a (poloidal)
magnetic field $\sim 10^{12}$, but the differential rotation winds it up
to a toroidal field of $\sim 10^{17}$G. In this case, $L_{em}$ is moderate
and the the main relativistic beam arises because of the annihilation of
such strong toroidal magnetic fields. However, following Duncan \&
Thompson (1992) and Usov (1992), even if we assume  the creation of a
super strong  magnetic field $B \sim 10^{15}$G in new born pulsars on the
plea that, unlike macroscopically turbulent accretion disks (actually
torus), the core of a pulsar is non-turbulent, even though hot, it is far
more difficult to conceive of a field $\sim 10^{17}$G in the absence of
real macroscopic quantum coherence. And in any case, because of ultrarapid
rapid gravitational energy loss, such models are unlikely to work.

In the NS-NS or NS-BH merger scenarios, for the resultant hot disk whose
temperature is higher than the superconductivity transition temperature
and which is macroscopically turbulent, it is extremely difficult to
conceive that it possesses a magnetic field higher than $B_q =4.4.
10^{13}$G. And note that, {\em in the post GRB971214 era}, one needs to
artificially further over stretch the value of $B_i$ to probably $10^{16}$G.

Could there be yet more novel variety of electromagnetic models of GRBs
involving compact objects? Recently, Usov (1998) purported to show that
the bare surface of a new born strange star would spontaneously emit pairs
with luminosity $L_\pm \sim 10^{51}$ erg/s for about $\sim 10$s. However,
we have already shown that this idea can not work simply because the time
scale to transport thermal energy from the core of the new born strange
star could be as large as $\sim 10^{13}$s, rather than $\sim 10$s, as
assumed by Usov. The only way to transport the thermal energy on the
desired short time scale of $\sim 10$s or shorter would be to consider the
emission of $\nu -\bar\nu$ pairs (Mitra 1998a).

As to the disk-jet case, the resultant value of $L_b$ would be much lower
than than the one estimated using vacuum electrodynamics because Ruffert
et al. (1997) show that the disk would be as hot, $T_{disk} \sim 10$MeV,
and therefore, even for an idealized thin disk geometry and no
quasi-spherical accretion, there would be fairly dense  plasma around the
disk (corona) and which would tend to quench any direct electromagnetic
mode.  Most probably, with {\em accretion rates exceeding the Eddington
rate by factors as large as} $10^{12}$, all such electromagnetic modes of
energy extraction, appropriate for radio pulsars, may not be applicable to
the GRB-disk at all.

\subsection{Mode of Energy Release}
 In general accretion energy will be channelized into X-rays and to
neutrinos along with the relativistic beam, if any. Given this fact, it is
not really possible  to ascertain how much of $E_{disk}$ would go into
$E_{beam}=L_b t_w$, where $t_w$ is the duration of the beam. Similarly,
the life time of the disk ($t_w$) also can not be predicted, and let us
just assume it to be $\sim 10$s.  Thus, the value of $L_b$ is to be set to
the desired value by keeping $B_i$ a free parameter and conveniently
ignoring other competetive and probably much more likely process of
neutrino production.

Yet, we can say that, as is the case for highly super-Eddington accretion,
accretion power will  primarily produce $\nu-\bar{\nu}$. In fact this is
the most natural process of cooling of hot and dense astrophysical plasma.
We may look at it in the following way.  One of the primary task assigned
by Nature to the electro-weak interaction is the drainage of energy from
physical and astrophysical systems. For macroscopic bodies  at relatively
lower densities and temperatures, it is the electromagnetic part which
undertakes the responsibility of relieving systems of excess energy and
help arrive at (new) equilibrium situations. On the other hand, at high
densities and temperatures, it is the weak processes which take up this
responsibility, and hot astrophysical bodies cool by emission of neutrinos
through pure electroweak processes (like at the late stages of evolution
of massive stars) or through various URCA type processes (Chiu \& Morrison
1960, Chiu \& Stabler 1961).

One may try to further appreciate this problem from this simple argument:
Even assuming that NS-NS or NS-BH collision process results in a disk of
necessary mass, {\em why must it get accreted within} $\sim 10s$
when the disk of Saturn can stay put indefinitely? Well, the answer would
be that there is viscous dissipation and associated loss of angular
momentum of the disk material. The nature and quantitative value of disk
viscosity is most uncertain and even if the source of viscosity is
considered to be disk magnetic field, the {\em dissipation of energy in
the disk necessarily means that accretion energy is primarily lost in the
form of heat and neutrinos}. Thus the BH-disk GRB model of Woosley (1993)
appropriately considers neutrino production as the main source of
accretion energy liberation. It is a different matter that, this model,
has difficulty in accounting for a value of $Q_{FB} >10^{50}$ erg.
Finally, note that, even in the AGN context, the physics of magnetically
dominated disk-jets is poorly understood and ``hydromagnetic propulsion as
a mechanism for accelerating jets has become attractive largely through a
process of elimination'' (Begelman 1994) of other competetive ideas.

It hardly requires a reminder that most of the other original cosmological
GRB models too logically envisaged that whether it is a NS-NS collision or
a NS-BH collision, the energy is liberated in the form of $\nu-\bar\nu$
(and in gravitational radiation) rather than in the form of any (direct)
relativistic $e^+ e^-$ beam (Goodman, Dar, \& Nussinov 1987, Paczynski
1990, Haensel, Paczynski, \& Amsterdamski 1991, Rees and Meszaros 1992,
Meszaros \& Rees 1992, Piran 1992, Mochkovitch et al. 1993).  Now falling
back on these works, we feel that, irrespective of the details, it is
much more likely that the powerful $e^+ e^- \gamma$ FB is indeed due to
the annihilation of $\nu +{\bar \nu} \rightarrow e^+ +e^-$. There may be
another point why we may be compelled to invoke neutrinos in this problem.

The sub-ms time structures found in many GRBs (Bhat et al. 1992) have
enhanced the general view that the dimension of the central engine of the
GRBs is $<10^7$cm.  Now we repeat the already oft-repeated argument that
the fact that for GRB971214, the luminosity is $L_\gamma \sim
10^{52}~\delta\Omega$ erg/s, with a probable  larger value of
$L_{FB}$, or for GRB970508, the value of $L_{FB} \sim 10^{52}  ~\delta
\Omega$ erg/s would suggest a temperature of the emission zone, $T_{em}
\sim 10^{10} -10^{11}$ K. At such high temperatures, even if one has no
neutrino to start with, photoneutrino processes (Chiu \& Morrison 1960)
ensure that neutrinos become much more numerous than $e^\pm$ pairs:
\begin{equation}
\gamma \rightarrow e^+ + e^- \rightarrow \nu_e +\bar{\nu_e}; \gamma +e^-
\rightarrow \gamma + \nu_e +\bar{\nu_e}
\end{equation}
The situation for such high temperature pair plasma becomes somewhat like
the early universe at corresponding temperature. In most of the realistic
astrophysical situations, neutrinos, being chargeless and weak particles,
have much better chance of escaping than $e^\pm$ pairs (the pair plasma is
already dominated by pairs at such temperatures).  Thus, unless we
conceive of highly fine tuned and  optimistic models, the very
existence of such extremely high values of $L_{FB}$ would imply that the
primary energy release mechanism from the central engine is in the form of
neutrinos.

In the context of the accretion induced collapse (of white dwarfs) model,
it has been estimated that,  the efficiency of the pair production via the
annihilation of $\nu -\bar\nu$ is
\begin{equation}
\eta_\pm \approx f \sigma_{\nu \bar\nu} {L_\nu \over \epsilon_\nu} {1\over
r_i c}
\end{equation}
Here the geometrical factor $f\sim 1$, $L_\nu $ is the luminosity of the
neutrino beam, $\sigma_{\nu \bar\nu} \sim 10^{-44}~ (\epsilon_\nu/10~{\rm
MeV})$ is the crosssection of the process averaged over three flavours,
and $\epsilon_\nu$ is the mean energy of the neutrinos (Goodman et al.
1987, Piran 1992). For the supernova case where $L_\nu
\sim 10^{52}$ erg/s and $\epsilon_\nu \sim 10-15$MeV, one obtains a very
low value of $\eta_\pm \sim 10^{-3}$.

If we directly and naively use this value of $\eta_\pm$ to understand the origin of
$Q_{FB} >  10^{53}$ erg, we would require a value of $Q_\nu \sim
10^{56}$ erg! But note here that, since $\sigma_{\nu \bar\nu} \propto
{\epsilon_\nu}^2$ we find
\begin{equation}
\eta_\pm \propto {\epsilon_\nu L_\nu \over r_i}
\end{equation}
 As $L_\nu$ increases beyond, say, $10^{52}$ erg/s, with a corresponding
decrease in $r_i$ (a deeper gravitational well necessary to have
incresed $L_\nu$), the value of  $\epsilon_\nu \sim (L_\nu /r_i^2)^{1/4}$
would also increase considerably. Therefore, eventually, we may have
\begin{equation}
\eta_\pm \propto  {L_\nu}^{1.25} r_i^{-1.5}
\end{equation}
By pursuing such probable  higher values of $L_\nu$,  one might partially
approach  a  situation, where the process $\nu +\bar\nu \rightarrow e^+
e^-$ is in thermal equilibrium with $\eta_\pm \sim 1$. It does not mean
that such a limit is exactly attained, but depending on the unknown
details of the evolution of the central engine, a value of $\eta_\pm
\rightarrow 0.01 -0.1$ may be easily achieved. For instance, a value of
$L_\nu \approx 10^{53}$ erg/s and $Q_\nu <10^{55}$ erg might
self-consistently explain the origin of $Q_{FB} >3. 10^{53}$ erg.

Here we note that some of the GRBs are stronger than even GRB971214. For
instance, the fluence of GRB980329 in the (50-300)KeV band is $5. 10^{-5}$
erg cm$^{-2}$, which is approximately five times larger than that of
GRB971214.  In fact the radio afterglow  has been detected for this burst
too (Taylor et al. 1998). If this burst too lie at $z >1$, it may be
possible to to postulate that the GRBs actually constitute a ``standard
candle'' probably within a factor of 10, with respect to the value of
$Q_\nu$.  Small variations in microphysics from one case to another may
induce a spread in the duration, $t_w$ (of the neutrino burst, not
necessarily same as the  duration of the observed GRB) with considerable
spread in the value of $L_\nu$. Then, because of the non-linear nature of
the conversion efficiency (eq. 30), there is additional spread, probably
spanning two orders of magnitude, in the eventual value of $Q_{FB}$. If
the baryon contamination is above a critical value (Shemi \& Piran 1990),
the resultant GRB could be very feeble  and most of the FB energy would go
into accelerating the baryons residing inside the FB and those lying ahead
(like previuosly ejected mass shells and presupernova wind). Such events
would be detectable in the electromagnetic band only if it occurs in a
nearby galaxy. It may be tempting to explain the association between the
weak event GRB 980425 ($Q_\gamma \sim 10^{48}$ ergs) with the mildly
relativistic extraordinary supernova event SN 1998bw having a kinetic
energy of $\sim 2.5. 10^{52}$ erg occuring at $z =0.0085$ (Galama et al.
1998, Soffita et al. 1998). We predict that the ejecta of SN 1998bw is not
highly beamed though it could be quasi spherical and quasi symmetric
contrary to several suggestions to this effect. Future observations should
confirm (or reject) this simple prediction.

Such high value of $Q_\nu >10^{54-54}$ erg, with an associated value of
$L_\nu \sim 10^{53}$ erg/s would demand a value of $\epsilon_\nu
\rightarrow 0.1$GeV. This difficult conclusions should not be avoided with
the plea that the existing models/theories are unable to explain the
origin of such prodigious neutrino luminosities. In fact the existing
paradigm, that gravitational collapse can not directly yield a value of
$Q_\nu$ larger than the BE of canonical NS, $Q_\nu \sim 3. 10^{53}$ erg, and
any attempt to harness more energy by studying collapse of relatively
more massive proto-NS would give birth to a black hole with hardly any
appreciable energy output {\em completely fails} to explain the genesis of
SN 1998bw. Of course, one can try to avoid this difficulty by proposing
``collapsar'' models (Woosley 1993, Woosley, Eastman, and Schmidt 1998).
Although, the intention of the present paper is not to outline any (new)
GRB model, we would only like to point out that, recently, we have shown
that, for continued general relativistic spherical collapse, the entire
origina mass energy can be radiated (Mitra 1998b,c).  But more
importantly, we have shown that general relativity actually inhibits the
formation of ``trapped surfaces'', the regions wherefrom even radiation
can not move out. This opens up the possibilty that gravitational collapse
of sufficiently massive stellar cores ($M_i > 3 M_\odot$) does not end up
as a quiet passage to a black hole.  Thus, the collapse of a $M_i<
10~M_\odot$ stellar core may account for this required $Q_\nu$. In fact,
we predict that, once the new generation giant neutrino detectors become
fully operational, it may be possible to detect neutrino bursts of value
$Q_\nu \sim 10^{54} -10^{55}$ erg with neutrino energies reaching $\sim 1$
GeV, in coincidence with the GRB events.

We are also aware here of the fact that, luminosity apart, for explaining
GRBs, one needs to conceive of situations where the baryon load of the
beam is modest and the value of $\Gamma$ is indeed high; but we would
address all such questions in a latter work.

To conclude, we did not attempt to predict the value of $Q_{FB}$ by
presuming some canonical value of $\xi_B$ and $\xi_e$. It is plausible
that these two parameters actually vary substantially from case to case
even for the afterglow regime. And, probably only when these parameters
have an appreciable value $\xi_B > 0.01$ and $\xi_e > 0.01$, radio
counterparts are found.  Conversely, the absence of detectable radio
afterglow for most of the GRBs may be ascribed to actual occurrence of
relatively lower vales of these parameters.

Whether the blast wave produes detectable radio afterglow or not, as
ensiaged previously (Mitra 1998d), there should be some remnants (GRBR),
like defunct supernova remnants, of these events. These remnants may be
found in nearby galaxies or in the Milky way.

\end{document}